\documentclass[aps,letterpaper,superscriptaddress,twocolumn,longbibliography,nofootinbib]{revtex4-1}
\pdfoutput=1

\usepackage{bmpsize}

\usepackage{hyperref}
\usepackage{color}
\usepackage{graphicx}	% Include figure filed
\usepackage[utf8]{inputenc} 
\usepackage{bm}
\usepackage{amsmath}
\usepackage{amsfonts}

\usepackage{eufrak}
\def\sumint{\,\hbox{$\sum$}\!\!\!\!\!\!\!\int}

\begin{document}

\title{Finite temperature CFT results for all couplings: O(N) model in 2+1 dimensions}

\author{Paul Romatschke}
\affiliation{Department of Physics, University of Colorado, Boulder, Colorado 80309, USA}
\affiliation{Center for Theory of Quantum Matter, University of Colorado, Boulder, Colorado 80309, USA}

\begin{abstract}
A famous example of gauge/gravity duality is the result that the entropy density of strongly coupled ${\cal N}=4$ SYM in four dimensions for large N is exactly 3/4 of the Stefan-Boltzmann limit. In this work, I revisit the massless O(N) model in 2+1 dimensions, which is analytically solvable at finite temperature $T$ for all couplings $\lambda$ in the large N limit. I find that the entropy density monotonically decreases from the Stefan-Boltzmann limit at $\lambda=0$ to exactly 4/5 of the Stefan-Boltzmann limit at $\lambda=\infty$. Calculating the retarded energy-momentum tensor correlator in the scalar channel at $\lambda=\infty$, I find that it has two logarithmic branch cuts originating at $\omega=\pm 4 T \ln \frac{1+\sqrt{5}}{2}$, but no singularities in the whole complex frequency plane. I show that the ratio 4/5 and the location of the branch points both are universal within a large class of bosonic CFTs in 2+1 dimensions.

\end{abstract}

\maketitle
%%%%%%%%%%%%%
%%%%%%%%%%%%%
\section{Introduction}
%%%%%%%%%%%%%
%%%%%%%%%%%%%
How does matter at finite temperature behave as the coupling is increased from very small to very high values? For very small couplings, traditional descriptions based on effective quasi-particles offer a very successful tool to calculate finite temperature matter properties (such as the entropy density $s$). However, at very large couplings, these descriptions break down because there no longer are well-defined quasi-particle degrees of freedom. For certain theories, such as ${\cal N}=4$ SYM in four dimensions in the limit of large number of colors, the strong coupling limit can be accessed using the gauge/gravity duality conjecture \cite{Maldacena:1997re}. Using calculations in accordance with this conjecture, it was found that the entropy density for infinite coupling $\lambda\rightarrow \infty$ was equal to exactly $3/4$ of the free-theory Stefan-Boltzmann value \cite{Gubser:1998nz},
\begin{equation}
  \label{eq:n=4}
  \lim_{\lambda\rightarrow \infty}s=\frac{3}{4} s_{\rm free}\,.
\end{equation}
Perturbative corrections to the free theory result and the infinite coupling result have been calculated (cf. Ref.~\cite{Gubser:1998nz}), but the full functional dependence of $\frac{s}{s_{\rm free}}$ on the coupling $\lambda$ for ${\cal N}=4$ SYM in four dimensions is not known (cf. the discussion in Ref.~\cite{Blaizot:2006tk}).

For general quantum field theories, the ratio $s/s_{\rm free}$ typically depends on both the coupling as well as the temperature $T$. However, for conformal field theories (CFTs) in $D+1$ space-time dimensions the entropy density is simply proportional to temperature $s\propto T^{D}$ because the trace of the energy-momentum tensor vanishes. Hence one may ask if there are examples of CFTs where one can evaluate the function $s/s_{\rm free}$ for \textit{all} values of the interaction strength. In the present work, I present the case of one such CFT where the ratio $s/s_{\rm free}$ can be calculated analytically for all couplings.

However, let me first discuss a model where $s/s_{\rm free}$ can be calculated analytically at infinite coupling. The model is the well-known O(N) model in the limit of $N\rightarrow \infty$ given by the Euclidean Lagrangian
\begin{equation}
  \label{eq:lag}
  {\cal L}=\frac{1}{2}\left(\partial_\mu \vec{\phi}\right)\cdot\left(\partial_\mu \vec{\phi}\right)+\frac{1}{2}m^2 \vec \phi^2+\frac{\lambda}{N}\left(\vec{\phi}^2\right)^2\,,
\end{equation}
in 3 Euclidean dimensions where $\vec{\phi}=\left(\phi_1,\phi_2,\ldots\phi_N\right)$. Here the Euclidean direction $x_0$ has been compactified on a circle of radius $\beta=\frac{1}{T}$, as is the standard procedure in thermal quantum field theory \cite{Laine:2016hma}. Note that in three dimensions, the coupling $\lambda$ has mass dimension one, and hence only the combination $\frac{\lambda}{T}$ is dimensionless in this theory. Thus, the high temperature regime corresponds to the weak coupling limit, whereas the low temperature regime corresponds to the strong coupling limit of the theory.

The O(N) model in 2+1d has been conjectured to be dual to higher spin gravity in AdS$_4$ in the large N limit \cite{Klebanov:2002ja}. Having access to the exact solution of the 2+1d O(N) model for large N, it is possible to evaluate finite-temperature real-time correlation functions. The analytic structure, and in particular the singularities of the retarded correlators would then correspond to the quasi-normal modes of the dual black brane geometry in AdS$_4$, providing further motivation for the present study.

It should be emphasized that several parts of the calculation presented here were already discussed in the literature more than 20 years ago, notably in Refs.~\cite{Sachdev:1993pr,Florkowski1994,Drummond:1997cw}. Perhaps because these calculations pre-date the advent of gauge/gravity correspondence, it seems that the available parts have not been assembled into a discussion of the exact finite-temperature properties of this CFT for all couplings, which is the subject of the present work.

\section{Calculation}

The partition function $Z=\int {\cal D}\phi e^{-\int {\cal L}}$ for the Lagrangian (\ref{eq:lag}) may be re-written by introducing
$$
1=\int {\cal D}\sigma \delta\left(\sigma-\vec{\phi}^2\right)=\int {\cal D}{\sigma}{\cal D}\zeta e^{i\int \zeta\left(\sigma-\vec{\phi}^2\right)}\,,
$$
performing the integral over $\sigma$, and separating the auxiliary field into a zero mode and fluctuations 
as 
\begin{equation}
  \label{eq:zgen}
  Z=\sqrt{\frac{\beta V N}{16 \lambda \pi}}\int d\zeta_0 e^{-\frac{\zeta_0^2 N \beta V}{16 \lambda}}\int {\cal D}\phi {\cal D}\zeta e^{-S_0-S_I}\,.
\end{equation}
Here $V$ is the volume of the Euclidean directions $x_1,x_2$. The action has been split into 
\begin{eqnarray}
  S_0&=&\frac{1}{2}\int d^3x\left[\left(\partial_\mu \vec{\phi}\right)\cdot\left(\partial_\mu \vec{\phi}\right)+m^2 \vec \phi^2+ i \zeta_0 \vec\phi^2+\frac{\zeta^{2} N}{2\lambda}\right]\,,\nonumber\\
  S_I&=&i \int d^3x \zeta \vec \phi^2\,,
\end{eqnarray}
and $\zeta_0$ denotes the global zero mode of $\zeta$. In the limit $N\rightarrow\infty$ one may neglect $S_I$, and the remaining path integrals can be performed analytically leading to
\begin{equation}
  \label{eq:part}
  Z= \sqrt{\frac{\beta V N}{16 \lambda \pi}}\int d\zeta_0 e^{-\beta V N \left(\frac{\zeta_0^2}{16 \lambda}+J\left(\sqrt{m^2+i \zeta_0}\right)\right)}\,,
\end{equation}
where denoting $\sumint_{\  K}=T \sum_n \int \frac{d^{2-2\epsilon}{\bf k}}{(2 \pi)^{2-2\epsilon}} $ in dimensional regularization gives \cite{Laine:2016hma}
\begin{equation}
  J(\alpha)=\frac{1}{2} \sumint_{K}\ln\left((2 \pi n T)^2+{\bf k}^2+\alpha^2\right)\,.
\end{equation}

The dimensionally regulated integral $J(\alpha)$ is finite for $\epsilon\rightarrow 0$, giving
%\begin{equation}
%  J(\alpha)=J_0+J_T\,, \quad J_0(\alpha)=-\frac{\alpha^{3}}{12\pi}\,,
%\end{equation}
%and
\begin{eqnarray}
  J(\alpha)&=&-\frac{T^3}{2\pi}\left[{\rm Li}_3\left(e^{\frac{\alpha}{T}}\right)-\frac{\alpha}{T} {\rm Li}_2\left(e^{\frac{\alpha}{T}}\right)+\frac{2\alpha^3}{6 T^3}\right.\nonumber\\
    &&\left.-\frac{\alpha^2}{2 T^2}\ln\frac{1-e^{\frac{\alpha}{T}}}{1-e^{-\frac{\alpha}{T}}}\right]\,.
  \end{eqnarray}

The property that the O(N) model has no logarithmic divergences in the large N limit is special to 2+1 dimensions. As a consequence, no renormalization is required, which allows setting the bare mass in the Lagrangian to zero, $m=0$. (Of course, radiative corrections requiring renormalization will arise once 1/N corrections are taken into account.)

In the large N limit, the partition function (\ref{eq:part}) can be evaluated exactly using the method of steepest descent, leading to
\begin{equation}
  \label{eq:partiition}
  \ln Z=\beta V N \left(-J\left(\sqrt{z^*}\right)+\frac{z^{* 2}}{16 \lambda}\right)\,,
\end{equation}
where $z^*=i\zeta_0$ is the stationary point of the action given as the solution of the equation
\begin{equation}
\label{eq:zgap}
  z^*=4 \lambda I\left(\sqrt{z^*}\right)\,,
\end{equation}
where
\begin{equation}
  I(\alpha)=2 \frac{d J(\alpha)}{d\alpha^2}=-\frac{\alpha}{4\pi}-\frac{T}{2\pi}\ln\left(1-e^{-\frac{\alpha}{T}}\right)\,.
\end{equation}
Putting $z^*=\xi^2 T^2$, Eq.~(\ref{eq:zgap}) becomes
\begin{equation}
  \label{eq:mygap}
  \frac{T}{\lambda}\xi^2=-\frac{\xi}{\pi}-\frac{2}{\pi}\ln\left(1-e^{-\xi}\right)\,,
  \end{equation}
which is similar to the result found in Ref.~\cite{Drummond:1997cw} for N=1 using self-consistent mean-field resummations. (See e.g. Ref.~\cite{Romatschke:2019wxc} for a discussion how the large N and mean-field resummation relate to each other in the context of scalar field theory.)

The finite-temperature pressure $P$ for the massless O(N) model in 2+1 dimensions in the large N limit is then found from (\ref{eq:partiition}) as
\begin{equation}
  P=T \frac{\partial \ln Z}{\partial V}=-N\left[J\left(\sqrt{z^*}\right)-\frac{z^{* 2}}{16 \lambda}\right]\,,
\end{equation}
and the entropy density is given by
\begin{eqnarray}
  \label{eq:entropy}
  s&=&\frac{\partial P}{\partial T}=-N \left.\frac{\partial J(\sqrt{z^*})}{\partial T}\right|_{z^*}\\
  s&=&\frac{N T^2}{4\pi}\left[\xi^3+\xi^2\ln \frac{1-e^{-\xi}}{(1-e^{\xi})^3}-6 \xi {\rm Li}_2\left(e^{\xi}\right)+6 {\rm Li}_3\left(e^\xi\right)\right]\,,\nonumber
  \end{eqnarray}
where in the first line the stationarity condition of the action (\ref{eq:zgap}) was used.

Another quantity of interest is the retarded correlation functions of the energy momentum tensor, e.g.
\begin{equation}
  G_R^{xy,xy}(\omega,{\bf p})=\int d^3x e^{i\omega t-i{\bf p}\cdot {\bf x}}\langle T^{xy}(x)T^{xy}(0)\rangle_R\,,
\end{equation}
where for simplicity I focus on vanishing external momentum ${\bf p}=0$ corresponding to the ``scalar channel'' in 2+1d in the following.  Note that the scalar channel correlator $G_R^{xy,xy}(\omega)$ possesses no hydrodynamic poles \cite{Baier:2007ix}. As a consequence, the physics of thermal widths that is required for hydrodynamic transport may be ignored for almost all frequencies, which allows me to neglect effects arising at subleading order in large N. (Note that thermal width may \textit{not} be ignored for frequencies $\frac{\omega}{T}\leq \frac{1}{N}$ because of the presence of so-called 'pinching poles' in the retarded correlator in this frequency regime \cite{Jeon:1994if}. Hence the following calculation of $G_R^{xy,xy}(\omega)$ can be expected to be accurate for all frequencies except those very close to the origin.)

The retarded correlator $G_R^{xy,xy}$ may be obtained by an analytic continuation of the Euclidean correlator as $G_R^{xy,xy}(\omega)=-\left.\langle T^{xy}T^{xy}\rangle_E(\omega_m)\right|_{\omega_m\rightarrow -i \omega+0^+}$ where 
\begin{equation}
  \langle T^{xy}T^{xy}\rangle_E(\omega_m)=2 N \sumint_K k_x^2 k_y^2 \Delta(\omega_n,{\bf k}) \Delta(\omega_n-\omega_m,{\bf k})\,,
  \end{equation}
with $\Delta(\omega_n,{\bf k})=\left[\omega_n^2+{\bf k}^2+z^*\right]^{-1}$ and where $z^*$ is the solution to Eq.~(\ref{eq:zgap}). Here $\omega_n=2\pi n T$ are the bosonic Matsubara frequencies. Carrying out the frequency sum and performing the angular average one finds
\begin{equation}
\label{eq:grres}
  G_R^{xy,xy}(\omega)=\frac{N}{4} \int \frac{dk k^{5-2\epsilon}}{(2\pi)^{1-2\epsilon}}\frac{\left(1+2 n(E_k)\right)E_k^{-1}}{( \omega+i0^+)^2-4 E_k^2}\,,
\end{equation}
where $n(x)=\left[e^{x/T}-1\right]^{-1}$ and $E_k=\sqrt{k^2+z^*}$. Writing $k^2=E_k^2-\frac{\omega^2}{4}-\left(z^*-\frac{\omega^2}{4}\right)$ and accordingly expanding the factor $k^4$ in the integrand, the retarded correlator is given by three parts, $G_R=G_1(\omega)+\left(z^*-\frac{\omega^2}{4}\right)G_2(\omega)+\left(z^*-\frac{\omega^2}{4}\right)^2G_3(\omega)$. The first two parts $G_1,G_2$ are easily evaluated using $n(E)=\sum_{n=1}^\infty e^{-n E/T}$, and do not contain logarithmic divergencies in dimensional regularization so that the limit $\epsilon\rightarrow 0$ may be taken. I find
\begin{eqnarray}
G_1(\omega)&=&-\frac{N T^3}{8\pi}\left[\frac{\xi \omega^2}{16 T^2}-\frac{\xi^3}{12}
+ {\rm Li}_3\left(e^{-\xi}\right)+\xi {\rm Li}_2\left(e^{-\xi}\right)\right.\nonumber\\
&&\left.
-\frac{1}{2}\left(\xi^2-\frac{\omega^2}{4 T^2}\right)\ln\left(1-e^{-\xi}\right)\right]\,,
\nonumber\\
G_2(\omega)&=&-\frac{N T}{16 \pi}\left[\xi+2 \ln\left(1-e^{-\xi}\right)\right]\,,\\
G_3(\omega)&=&\frac{N}{4}\int_{\xi T}^\infty \frac{d E}{2\pi} \frac{1+2 n(E)}{(\omega+i 0^+)^2-4 E^2}\,,\nonumber
\end{eqnarray}
where again $z^*=\xi^2 T^2$ was used. The part $G_3$ does not seem to be expressible in closed form in terms of elementary functions. However, the analytic structure of $G_3$ may be gleaned from writing $1+2 n(E)=T \sum_{n=-\infty}^\infty \frac{2 E}{\omega_n^2+E^2}$ and performing the integral over $E$, finding
\begin{equation}
\label{eq:g3}
G_3(\omega)=\frac{N}{8 \pi T}\sum_n \frac{\ln\frac{\xi^2-\frac{\omega^2}{4 T^2}}{\xi^2+(2 \pi n)^2}}{\frac{\omega^2}{T^2}+4 (2 \pi n)^2}\,.
\end{equation}

\section{Results}

As outlined in the introduction, because the only dimensionless parameter for the theory is $\frac{T}{\lambda}$, the weak coupling limit corresponds to the limit of $T\rightarrow \infty$. In this case, Eq.~(\ref{eq:mygap}) gives $\xi^2\rightarrow \frac{\lambda}{\pi T}W\left(\frac{\pi T}{\lambda}\right)\rightarrow 0$ where $W(x)$ is the product log function. It is straightforward to find the entropy density (\ref{eq:entropy}) for $\xi\rightarrow 0$ in this case:
\begin{equation}
  \label{eq:sfree}
\lim_{\lambda/T\rightarrow 0}s=\frac{3 N T^2 \zeta(3)}{2 \pi}=s_{\rm free}\,.
\end{equation}
In the opposite limit of strong coupling, the solution to (\ref{eq:mygap}) is $\xi\rightarrow 2\ln \frac{1+\sqrt{5}}{2}$ for which Eq.~(\ref{eq:entropy}) gives
\begin{equation}
  \label{eq:ssrong}
\lim_{\lambda/T\rightarrow \infty}s=\frac{12 N T^2 \zeta(3)}{10 \pi}=\frac{4}{5}s_{\rm free}\,,
\end{equation}
which was pointed out already in Refs.~\cite{Sachdev:1993pr,Drummond:1997cw}. One may ask if the theory is a CFT in these two regimes. To this end, one calculates the trace of the energy momentum tensor in three dimensions as
\begin{equation}
  T^{\mu}_\mu=s T-3 P=\frac{\xi^4 T^4 N}{16 \lambda}\,,
\end{equation}
where (\ref{eq:mygap}) has been used. Since $\xi\rightarrow 2\ln \frac{1+\sqrt{5}}{2}$ one has $\lim_{\lambda/T\rightarrow \infty} T^\mu_\mu=0$ and the theory is a CFT for infinite and vanishing coupling, but not in-between.

%For intermediate values of $\frac{\lambda}{T}$, $s/s_{\rm free}$ can be evaluated numerically from Eq.~(\ref{eq:entropy}). The corresponding result is shown in Fig.\ref{fig1}. I find that $s/s_{\rm free}$ decreases monotonically with increasing coupling constant.

Inspecting the retarded correlator in the scalar channel (\ref{eq:grres}), one finds that for non-zero coupling (and hence $\xi\neq 0$) the parts $G_1$ and $G_2$ are analytic in the whole complex frequency plane, while $G_3$ has logarithmic branch cuts originating from the points $\omega=\pm 2 \xi T$, reminiscent of the weak-coupling results found in Ref.~\cite{Florkowski1994}. However, note that $G^{xy,xy}_R(\omega)$ does not have any singularities or poles. The logarithmic singularities at $\omega=\pm 2 \xi T$ are canceled by the factor $\xi^2-\frac{\omega^2}{4 T^2}$ multiplying $G_3$ and the limit $\omega\rightarrow 2 i\omega_n$ of $G_3$ is regular.
(It should be pointed out that in the case of 3+1d and weak coupling $\xi\rightarrow 0$, $G_3$ possesses poles at $\omega=-4\pi i T n$, $n\in \mathbb{N}$ similar to the weak-coupling results for the glueball correlator in Ref.~\cite{Hartnoll:2005ju}.)

The singularity structure of $G_3(\omega)$ consisting of two logarithmic branch cuts does not change qualitatively when changing the coupling $\lambda$ as long as $\frac{\lambda}{T} \gg \frac{1}{\rm N}$. The origin of the branch cuts is governed by the solution of Eq.~(\ref{eq:mygap}), ranging from $\frac{\omega}{T}=\pm 4 \ln \frac{1+\sqrt{5}}{2}\simeq 1.92$ at $\lambda=\infty$ to $\omega\rightarrow 0$ for $\lambda\rightarrow 0$. Note that the result at $\lambda=\infty$ matches the structure found for the charge response found in Ref.~\cite{WitczakKrempa:2012gn}.

However, when $\frac{\lambda}{T} \sim \frac{1}{\rm N}$ or smaller, the structure of $G_R(\omega)$ near $\omega=0$ is sensitive to 1/N effects such as the thermal widths. Since these effects are not included in the calculation of Eq.~(\ref{eq:g3}), the singularity structure of $G_R(\omega)$ may differ from those discussed above when $\frac{\lambda}{T} \ll \frac{1}{\rm N}$.

\begin{figure*}[t]
  \includegraphics[width=0.7\linewidth]{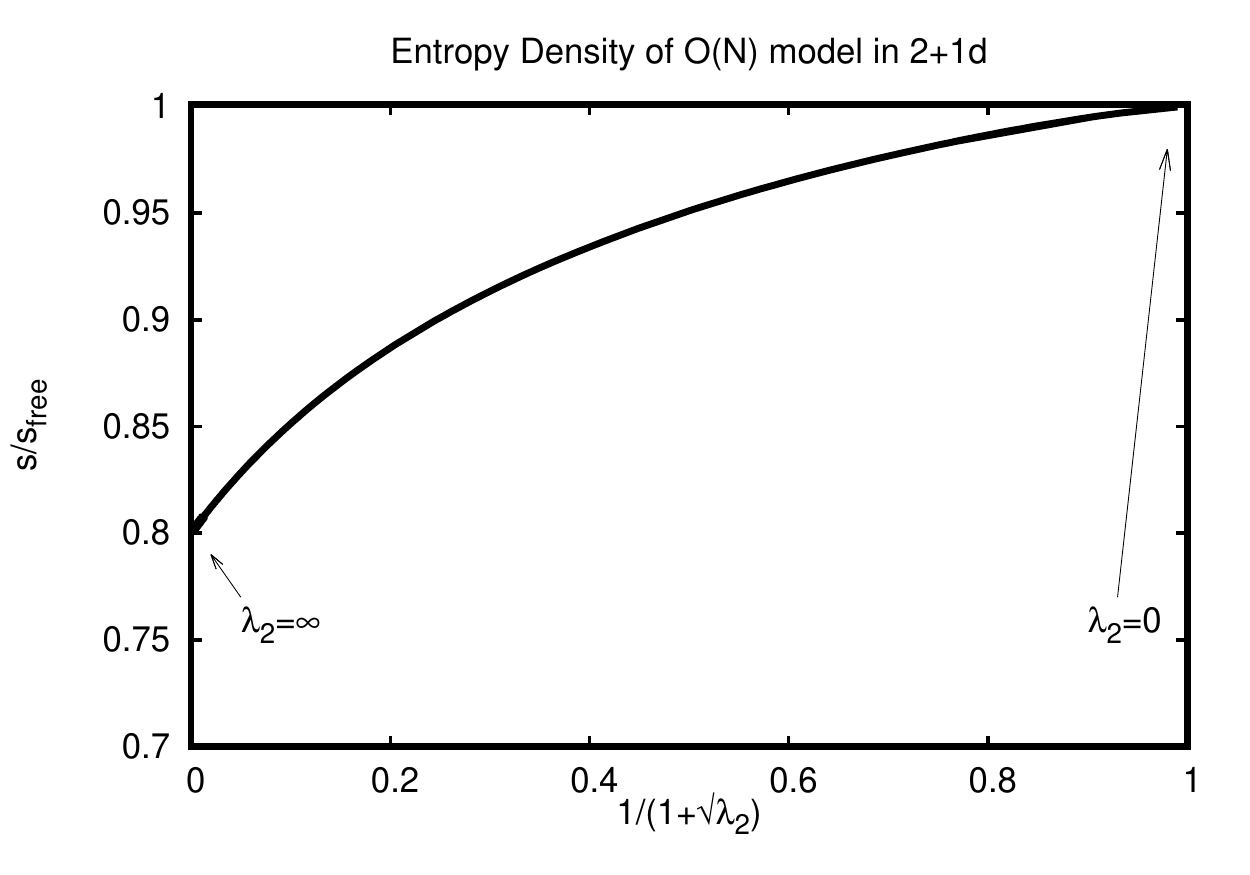}
  \caption{\label{fig1} The ratio $\frac{s}{s_{\rm free}}$ for the O(N) model with sextic interactions in 2+1 dimensions in the large N limit for all temperatures/couplings. Results are shown using a compactified interval $\frac{1}{1+\sqrt{\lambda_2}}\in [0,1]$ in order to show all coupling values. Arrows indicate free theory result (\ref{eq:sfree}) and strong coupling result (\ref{eq:ssrong}). See text for details.}
\end{figure*}

\section{Universality in the Strong Coupling Limit} 
\label{sec:u}

The factor of 4/5 for the O(N) model in Eq.~(\ref{eq:ssrong}) in 2+1 dimensions is suspiciously similar to the factor 3/4 found for ${\cal N}=4$ SYM in 3+1 dimensions. As a result, one might wonder if the similarity between a purely bosonic theory in 3 dimensions and a supersymmetric theory in 4 dimensions is just a fluke. To address this issue, I calculate the ratio $s/s_{\rm free}$ at zero temperature/infinite coupling in a large class of stable bosonic large N CFT in 2+1 dimensions. Let me thus consider a Lagrangian density where the potential part of (\ref{eq:lag}) is modified to
\begin{equation}
\frac{\lambda}{N}\left(\vec{\phi}^2\right)^2\rightarrow N\times U\left(\vec{\phi}^2/N\right)\,,
\end{equation}
with $U\left(x\right)$ an arbitrary function that should possess no local minima except for $U^\prime(0)=0$ with $U^{\prime\prime}(0)>0$ (to avoid spontaneous mass generation and hence breaking the CFT) and fulfill $U(x\rightarrow \infty)\rightarrow \infty$. Introducing the auxiliary fields $\sigma,\zeta$ as before, the large N partition function is given by 
\begin{equation}
\ln Z=\beta V N \left(- J\left(\sqrt{z^*}\right)+\frac{z^* \sigma^*}{2}-U(\sigma^*)\right)\,,
\end{equation}
with $z^*,\sigma^*$ given by the stationarity condition of the action
\begin{equation}
  \label{mygappies}
  \sigma^*=I(\sqrt{z^*})\,,\quad z^*=2 U^\prime(\sigma^*)\,.
\end{equation}
(Note that for $U(\sigma)=\lambda \sigma^2$ these reduce to the gap equation (\ref{eq:zgap})). In a CFT, there are no zero temperature mass scales, so by dimensional reasons $z^*=\xi^2 T^2$. In the low temperature limit then $z^*\rightarrow 0$ and hence $U^\prime (\sigma^*)\rightarrow 0$. Since our condition on $U(\sigma)$ was that its only minimum was at $\sigma=0$, we can expand $U^\prime(\sigma)$ around $\sigma=0$ as
\begin{equation}
  \label{eq:ggap2}
  U^\prime(\sigma^*)=2 \lambda_1 \sigma^*+3 \lambda_2 \sigma^{* 2}+4 \lambda_3 \sigma^{* 3}+\ldots=\frac{\xi^2 T^2}{2}\,,
\end{equation}
where $\lambda_1,\lambda_2,\lambda_3,\ldots$ are coupling constants. The constraint $U^{\prime\prime}(0)>0$ implies that $\lambda_1\geq 0$ and hence the solution to (\ref{eq:ggap2}) for $T\rightarrow 0$ is $\sigma^*=\frac{\xi^2 T^2}{4 \lambda_1}$. Plugging this value for $\sigma^*$ into (\ref{mygappies}) leads to
\begin{equation}
  \frac{\xi^2 T}{4 \lambda_1}=\frac{I(\xi T)}{T}=-\frac{\xi}{\pi}-\frac{2}{\pi}\ln\left(1-e^{-\xi}\right)\,.
\end{equation}
In the zero-temperature limit, the left-hand side of this gap equation vanishes, and one finds $\xi=2\ln\frac{1+\sqrt{5}}{2}$, as before. This shows that the thermal mass in the zero temperature limit is always given by twice the logarithm of the golden ratio, regardless of the form of the potential $U(\sigma)$ within the class considered here. As a consequence, the branch points of the retarded correlator $G^{xy,xy}_R(\omega)$ located at $\frac{\omega}{T}=\pm 4 \ln\frac{1+\sqrt{5}}{2}$ can be regarded as universal in the low temperature/strong coupling limit.

A similar feature holds for the entropy density. Since the implicit temperature dependence in $z^*,\sigma^*$ does not contribute to the temperature derivative in (\ref{eq:entropy}) because of the stationarity condition (\ref{mygappies}), the result found in Eq.~(\ref{eq:entropy}) is universal. As a consequence,
\begin{equation}
  \label{eq:main}
\frac{s}{s_{\rm free}}=4/5\,,
\end{equation}
holds for all large class of bosonic large N CFTs in 2+1 dimensions in the strong coupling limit.

\section{A CFT for all couplings: sextic interactions}

A potential $U$ of particular interest is the case of sextic interactions, or
\begin{equation}
  U(\sigma)=\lambda_2 \sigma^3\,,
\end{equation}
where unlike the quartic interaction case considered above, $\lambda_2$ is dimensionless. The discussion from section \ref{sec:u} is modified by considering $\lambda_1=0$, finding
\begin{equation}
  \label{sexticgap}
  \frac{4 \xi}{\sqrt{6 \lambda_2}}=-\frac{\xi}{\pi}-\frac{2}{\pi}\ln\left(1-e^{-\xi}\right)\,,
  \end{equation}
for the location of the saddle point $z^*=\xi^2 T^2$, and
\begin{equation}
  P=-N\left[J(\sqrt{z^*})-\frac{2}{\sqrt{\lambda_2}} \left(\frac{z^*}{6}\right)^{3/2}\right]\,,
\end{equation}
for the pressure of the theory. The entropy density is still given by (\ref{eq:entropy}) such that the trace of the energy momentum tensor becomes
\begin{eqnarray}
  T^\mu_\mu&=&s T-3 P=\frac{\xi^2 T^3 N}{4}\left(\frac{\xi}{\pi}+\frac{2}{\pi}\ln\left(1-e^{-\xi}\right)+\frac{4 \xi}{\sqrt{6\lambda_2}}\right)\,,\nonumber\\
  &=&0\quad \forall \lambda_2\,,
\end{eqnarray}
where (\ref{sexticgap}) has been used. Therefore, the massless O(N) model with sextic interaction is a CFT for all values of the coupling $\lambda_2$, and fulfills (\ref{eq:main}) as before. For intermediate values of $\lambda_2$, $s/s_{\rm free}$ can be evaluated numerically from Eq.~(\ref{eq:entropy}). The corresponding result is shown in Fig.\ref{fig1}. I find that $s/s_{\rm free}$ decreases monotonically with increasing coupling constant.

  \section{Discussion}

  In this work, exact results for the O(N) model in 2+1 dimensions at finite temperature  in the large N limit have been discussed. It was found that the O(N) model has a weak-strong relation of the entropy density similar to the famous 3/4 factor for ${\cal N}=4$ SYM in Eq.~(\ref{eq:n=4}), except that the ratio happens to be 4/5 instead of 3/4, and that the calculation was done entirely on the field-theory side without any invocation of gauge/gravity duality. For the O(N) model, the factor of 4/5 is simply the result of a finite thermal mass for $\lambda\rightarrow \infty$, leading to an effective reduction in the number of degrees of freedom at strong coupling compared to weak coupling, and the result is universal within a large class of bosonic CFTs.

  Whereas for ${\cal N}=4$ SYM only perturbative expansions for the ratio $s/s_{\rm free}$ at $\lambda=0$ and $\lambda=\infty$ exist, the result (\ref{eq:entropy}) presented here for the O(N) model with sextic interactions is valid for all couplings. To my knowledge, this is only example of a finite-temperature CFT where the entropy density is known for arbitrary coupling.

In addition to the entropy density, the retarded energy-momentum tensor correlator  in the scalar channel was calculated at finite temperature. This quantity is of interest in the context of gauge/gravity duality because the singularity structure of $\langle T^{xy} T^{xy}\rangle_R$ is expected to correspond to the quasi-normal mode frequencies of a black-brane in AdS$_4$ \cite{Son:2002sd} with Hawking temperature $T$. It was found that the retarded correlator in the scalar channel in the large N, large coupling limit possesses logarithmic branch cuts originating from the points $\omega=\pm 4 T \ln \frac{1+\sqrt{5}}{2}$, but no singularities. The structure of the retarded correlator stays qualitatively the same as the coupling is decreased from infinity, but the origin of the branch cuts move closer to the origin.

Also, given the conjectured duality of the O(N) model in the large N limit, the present results may allow for the possibility of studying alternative theories of gravity in a controlled setting.

It is likely that the present calculation can be extended to include 1/N effects, e.g. using the non-perturbative resummations schemes presented in Refs.~\cite{Romatschke:2019rjk,Romatschke:2019wxc}. It is my hope that this could help answer a variety of open questions, such as transport phenomena in quantum field theories at arbitrary coupling.

  \section{Acknowledgments}

  This work was supported by the Department of Energy, DOE award No DE-SC0017905. I would like to thank Poul Damgaard, Tom DeGrand, Steve Gubser, Mikko Laine, William Witczak-Krempa, Oliver DeWolfe and Bill Zajc for helpful discussions.

\bibliography{cft}

\end{document}